\documentclass[a4paper,11pt]{jpconf}
\usepackage{graphicx}
\newcommand{\be}{\begin{equation}}
\newcommand{\ee}{\end{equation}}
\newcommand{\bra}{\langle}
\newcommand{\ket}{\rangle}
\newcommand{\bea}{\begin{eqnarray}}
\newcommand{\eea}{\end{eqnarray}}

\begin{document}
\title{Analysis of Spin Financial Market by GARCH Model}

\author{Tetsuya Takaishi}

\address{Hiroshima University of Economics, Hiroshima 731-0192, JAPAN}

\ead{tt-taka@hue.ac.jp}

\begin{abstract}
A spin model is used for simulations of financial markets.
To determine return volatility in the spin financial market 
we use the GARCH model often used for volatility estimation in empirical finance. 
We apply the Bayesian inference performed by the Markov Chain Monte Carlo method
to the parameter estimation of the GARCH model.
It is found that volatility determined by the GARCH model exhibits 
"volatility clustering" also observed in the real financial markets.
Using volatility determined by the GARCH model we examine the mixture-of-distribution hypothesis (MDH) 
suggested for the asset return dynamics.
We find that the returns standardized by volatility 
are approximately standard normal random variables.
Moreover we find that the absolute standardized returns
show no significant autocorrelation.
These findings are consistent with the view of the MDH for the return dynamics.

\end{abstract}

\section{Introduction}
Statistical properties of asset price returns
have been extensively  studied and some
pronounced properties are found and classified as stylized facts\cite{CONT}. 
One of the stylized facts known for many years is 
that the probability distributions of returns exhibit
fat-tailed distributions. 
This evidence indicates that asset price dynamics is not a simple Gaussian random walk.
A possible origin for the fat-tailed distributions has been explained 
by the mixture-of-distribution hypothesis (MDH)\cite{Clark} where 
the price return dynamics is described by 
a Gaussian random process with time-varying volatility.
Let $r(t)$ be a return at time $t$.
Then under the MDH the return is described by
$r(t)=\sigma_t \epsilon_t$, where $\sigma_t$ is a standard deviation 
and $\epsilon_t$ is a Gaussian random number with variance 1 and mean 0. 
In finance $\sigma_t^2$ is also called "volatility".
Using high-frequency financial data  
the MDH has been examined and the evidence on the MDH 
has been found\cite{Andersen1,Andersen2,Andersen3,Andersen4,Fleming,TakaishiRV,TakaishiRV2}.

Empirically it is well known that 
the volatility changes with time and exhibits
persistence of the same magnitude of the volatility.
This feature of the volatility is called "volatility clustering"
which is also one of the stylized facts.
In order to forecast future volatility one need to use 
volatility models which possess the properties of return and volatility.  
The most successful models in empirical finance 
are the ARCH model\cite{ARCH} and its generalized version, the GARCH model\cite{GARCH}.
Since the invention of the ARCH and GARCH models 
many extended models have been also proposed and applied for empirical finance.
Some examples of those models are EGARCH\cite{EGARCH}, QGARCH\cite{QGARCH1,QGARCH2}, 
GJR\cite{GJR}, APARCH\cite{APARCH} and GARCH-RE\cite{GARCHRE} models etc. See also, e.g. \cite{Bollerslev}.

In physics to understand financial market dynamics a variety of agent-based models 
have been proposed and examined\cite{Bouchaud,Stauffer,Sato,Lux,Iori,Stauffer2,AGENT,Bornholdt,
SPIN1,Sanchez,Yamano,SPIN5,TakaishiSP,Bornholdt2}.
It is found that to some extent those models  are able to capture some of stylized facts 
such as fat-tailed distributions of returns
and long autocorrelation times of the absolute returns.

In this study we perform simulations of financial markets in a three-state spin model\cite{TakaishiSP} which is 
an extended version of the two-state Bornholdt model\cite{Bornholdt} and determine volatility of returns simulated in the model. 
In the real financial markets volatility itself is not directly observed.
Therefore in order to infer  the latent volatility one usually utilizes parametric models  such as the GARCH model.
Here we also take the same approach and use the GARCH model to determine the spin-market volatility.
After determining the volatility 
we further examine the view of the MDH for the return dynamics of the financial spin model.

\section{Three-state spin model}
The model we use here is a Potts-like model\cite{TakaishiSP} in which 
agents ( or spins $S_i$ ) locate on one of lattice sites, having one of three states $(1,-1,0)$. 
The three states $(1,-1,0)$ are assigned to "buy", "sell" and "inactive" orders respectively.
The model includes two interactions which conflict each other.
One is the nearest neighbor interaction which causes the ferromagnetic order 
or the majority effect.
In other words  with this interaction the agents tend to imitate their neighbor agents
and as a result belong to the majority group.
In the real financial markets this imitation corresponds to the herding behavior. 
The other interaction is a global interaction proportional to 
the magnitude of the magnetization, which causes the anti-ferromagnetic order 
or the minority effect. 
When the magnitude of the magnetization is big, most of spins takes 1 ( or -1 ).
Such a state corresponds to "bubble" state in the real financial markets.
When the bubble state appears agents should change their spins to the minority group.
By doing so agents are able to avoid future possible loss when the "bubble" economy bursts.
These two interactions conflicting each other cause a complicated dynamics. 
For instance we observe no single stable phase,
but instead find ferromagnetic and anti-ferromagnetic phases repeatedly.
We again come to this point below.

Based on the heat bath dynamics proposed by Bornholdt\cite{Bornholdt} 
we update spins according to the following probability.
\be
P(S_i \rightarrow S_i^{\prime})= \left\{
\begin{array}{ll}
\exp \left( \lambda(h(i,S_i^{\prime})-\mu S_i  S_i^\prime|M|) \right)/C & \mbox{   for   }   S_i^\prime=1,-1 \\
\\
\exp \left( \lambda(h(i,S_i^{\prime})-\mu(\epsilon_i |M| -2\gamma K)) \right)/C  & \mbox{   for   }   S_i^\prime= 0
\end{array}
\right.
\label{eq:gamma}
\ee 
where
\be
h(i,S_i^{\prime}) =\sum_{\bra i,j \ket} \delta_{S_j,S_i^{\prime}},
\ee
$\bra i,j \ket$ stands for the summation over the nearest neighbors $j$ of the site $i$
and $\epsilon_i$ is defined by $\epsilon_i=2\delta_{S_i,0}-1$ which takes 1 for $S_i =0 $ and
$-1$ for $S_i \neq  0$.
$C$ is the normalization factor determined so that the following equation is satisfied.
\be
\sum_{k=1,-1,0}P(S_i \rightarrow k)=1 .
\ee
$M$ stands for the magnetization calculated by
\be
M=\frac1N \sum_{i=1}^N S_i,
\label{eq:M}
\ee
where $N$ is the number of agents in the system 
and $K$ is the inactivity rate given by
\be
K=\frac1N \sum_{i=1}^N \delta_{S_i,0}.
\ee

The Hamiltonian of the model could be written as 
\be
H=-J \sum_i h(i,S_i) + \sum_i L_i,
\ee
where
\be
L_i = \left\{
\begin{array}{ll}
\mu {\rm sign}(S_i) S_i |M|  & \mbox{   for   }   S_i=1,-1 \\ 
\\
\mu(\epsilon_i |M| -2\gamma K)   & \mbox{   for   }   S_i= 0
\end{array}
\right.
\ee
and $J=1$ for a ferromagnetic coupling.
The partition function $Z(\lambda,H)$ is given by
\be
Z(\lambda,H)= \sum_{S_1,...,S_N} \exp(-\lambda H).
\ee

To make an insight on the model
let us consider two-state model (Ising-type model) for simplicity.
For the two-state model the Hamiltonian can be written as
\be
H=-J \sum_i h(i,S_i) + \sum_i \mu {\rm sign} (S_i) S_i |M| .
\ee
With a redefined coupling $J=2J^\prime$ the nearest neighbor interaction term 
may be rewritten as
\be
H=- J^\prime \sum_i 2(h(i,S_i)-2d) + \sum_i \mu {\rm sign}(S_i) S_i |M|,
\label{eq:2dpotts}
\ee
where $d$ is the dimension of the model, i.e. $d=2$ in this case.
Equation (\ref{eq:2dpotts}) corresponds to the usual Ising model representation as 
\be
H=-J \sum_{\bra i,j \ket} S_j S_i  + \sum_i \mu {\rm sign}(S_i) S_i |M|.
\ee
Let $E_l$ be a term containing $S_l$ and we obtain
\be
E_l=
-(J \sum_{\bra l,j \ket} S_j - \mu {\rm sign}(S_l) |M|)S_l.
\ee
Using the mean field $\phi=\bra S_l \ket$ we obtain 
\be
E_l=
-(J z \phi  - \mu {\rm sign}(S_l) |M|) S_l,
\ee
where $z=2d$ is the number of the neighbor sites of an agent $S_l$.
Each agent may interact with some other agents.
In the realistic situation the number of agents from which an agent can obtain information on the markets
should be limited. Therefore  a very big $z$ for which the mean field approximation is appropriate may not be realistic and, 
certainly the infinite $z$ is not. There also exists a warning that  the mean field approximation should not be used for
agent models\cite{MFA}.
Since we already know that simulations on $d=2$ \cite{Bornholdt,SPIN5} and on low dimensions\cite{Yamano}
can successfully produce some similarity with the real financial markets such as the power-law return distribution,
here we employ 2 dimensional lattice simulations. 

Contrast to the original Ising model, 
a term $\mu {\rm sign}(S_l) |M|$  behaves differently, i.e.
it changes the {\rm sign} depending on $S_l$ and $|M|$ varies with time.
These features cause a more complex behavior.
Setting the temperature $T$  $(\sim 1/\lambda)$ below the critical temperature $T_c$ and $\mu=0$,
the system of this model is in the ordered phase. 
In addition, assuming $|M|$ is a constant,
Reference \cite{Bornholdt2} considered the phase diagram as a function of $|M|$ 
and found that the system undergoes the phase transition to the disordered phase at some critical $|M|_c$.
In the disordered phase, $M$ fluctuates strongly, that creates a bigger volatility.
On the other hand in the ordered phase the fluctuation of $M$ is small, 
which results in exhibiting a small volatility.
When we let $|M|$ vary, the system can switch the phase to the other one repeatedly in simulations.
The durations of the phases correspond to the appearance of the volatility clustering.

The introduction of $\gamma$ in (\ref{eq:gamma}) gives a more flexibility to the model. 
By varying $\gamma$ one can observe the exponential return distribution\cite{TakaishiSP}, in addition to the power-law distribution.
Empirically the  exponential return distributions have been also observed in the Indian market\cite{Indian}.
Since $\gamma$ can change the number of agents who participate buy-or-sell transactions,
$\gamma$ might be related to tune the number of transactions or volume.

\section{GARCH Model}

The GARCH(p,q) model by Bollerslev\cite{GARCH} is given by
\be
y_t=\sigma_t \epsilon_t ,
\ee
\be
\sigma_t^2  = \omega + \sum_{i=1}^{q}\alpha_i y_{t-i}^2
+ \sum_{i=1}^{p}\beta_i \sigma_{t-i}^2,
\ee
where the GARCH parameters are restricted to $\omega>0$, $\alpha_i>0$ and $\beta_i>0$ to ensure a positive volatility.
$\epsilon_t$ is an independent normal error $\sim N(0,1)$ and returns are given by $y_t$.

We focus on the GARCH(1,1) model
where the volatility process is given by
\be
\sigma_t^2  = \omega + \alpha y_{t-1}^2 + \beta \sigma_{t-1}^2.
\label{eq:GARCH11}
\ee
Although the GARCH(1,1) model is the simplest,  in empirical studies
the GARCH(1,1) model is often chosen as the best one by comparison 
of information criterions such as AIC\cite{AIC}.  
In this study the GARCH(1,1) model is denoted simply by GARCH model.

The unconditional volatility $\bar{\sigma}^2$ can be found 
by substituting $\sigma_t^2=\sigma_{t-1}^2=\bar{\sigma}^2$ and
$E[y_{t-1}^2]=\sigma_t^2$ to (\ref{eq:GARCH11}).
For $(\alpha+\beta)<1$, we obtain
\be
\bar{\sigma}^2=\frac{\omega}{1-(\alpha+\beta)}.
\ee
The rate of convergence to the unconditional volatility
can be measured by $\alpha+\beta$.
When  $\alpha+\beta$ is near 1, volatility persists 
very long. Empirically the value of $\alpha+\beta$ is often inferred 
to be close to 1. 

The GARCH model includes three model parameters, $\alpha,\beta$ and $\omega$.
These parameters are determined so that the model matches the observed time series data.
To determine the model parameters we employ the Bayesian inference performed by
the Markov Chain Monte Carlo (MCMC) method.
The MCMC method we use here is the Metropolis-Hastings method with
an adaptive multi-dimensional student's t-distributions, which is shown to 
be very efficient for the Bayesian inference of the GARCH model\cite{Takaishi1,Takaishi2,Takaishi3,Takaishi4}.

\begin{figure}[ht]
\vspace{1cm}
%\begin{miniprre}{14pc}
\begin{center}
\includegraphics[width=17pc]{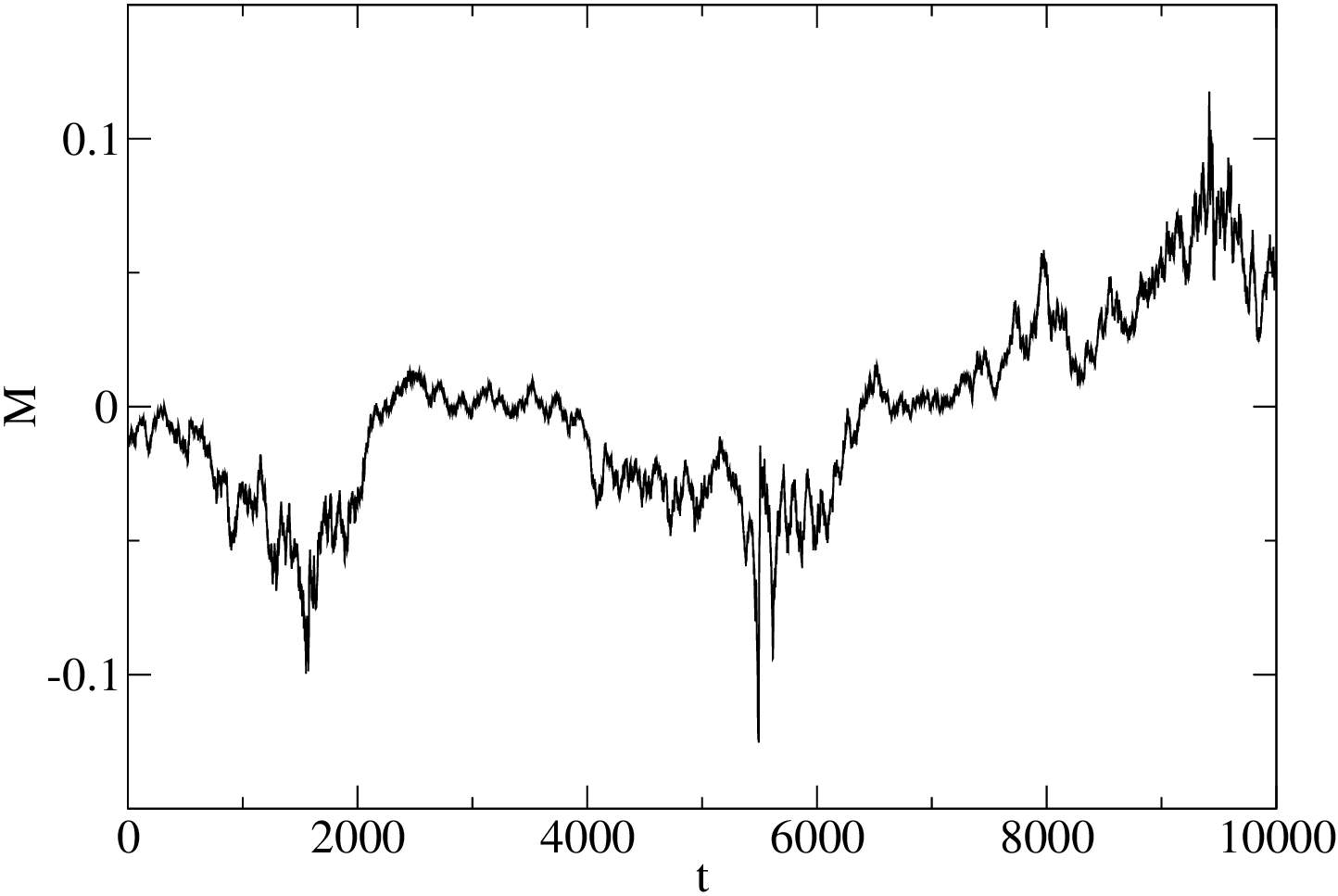}
\vspace{5mm}
\includegraphics[width=17pc]{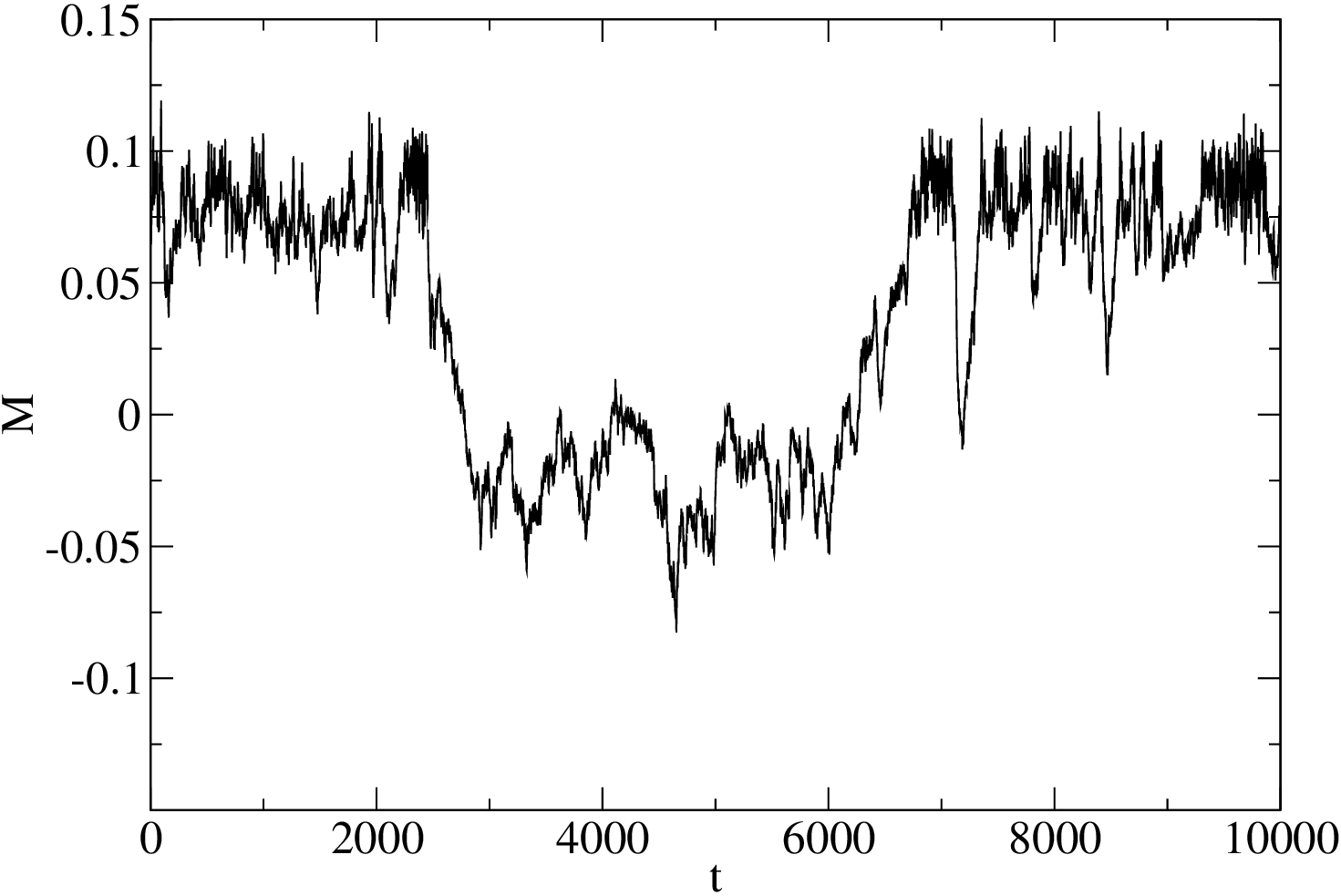}
\end{center}
\caption{
Magnetization $M(t)$ simulated at $\gamma=1.0$ (left) and $\gamma=0.05$ (right). }
%\end{minipage}
%\hspace{5pc}%
\end{figure}

\begin{figure}[ht]
\vspace{1cm}
%\begin{miniprre}{14pc}
\begin{center}
\includegraphics[width=17pc]{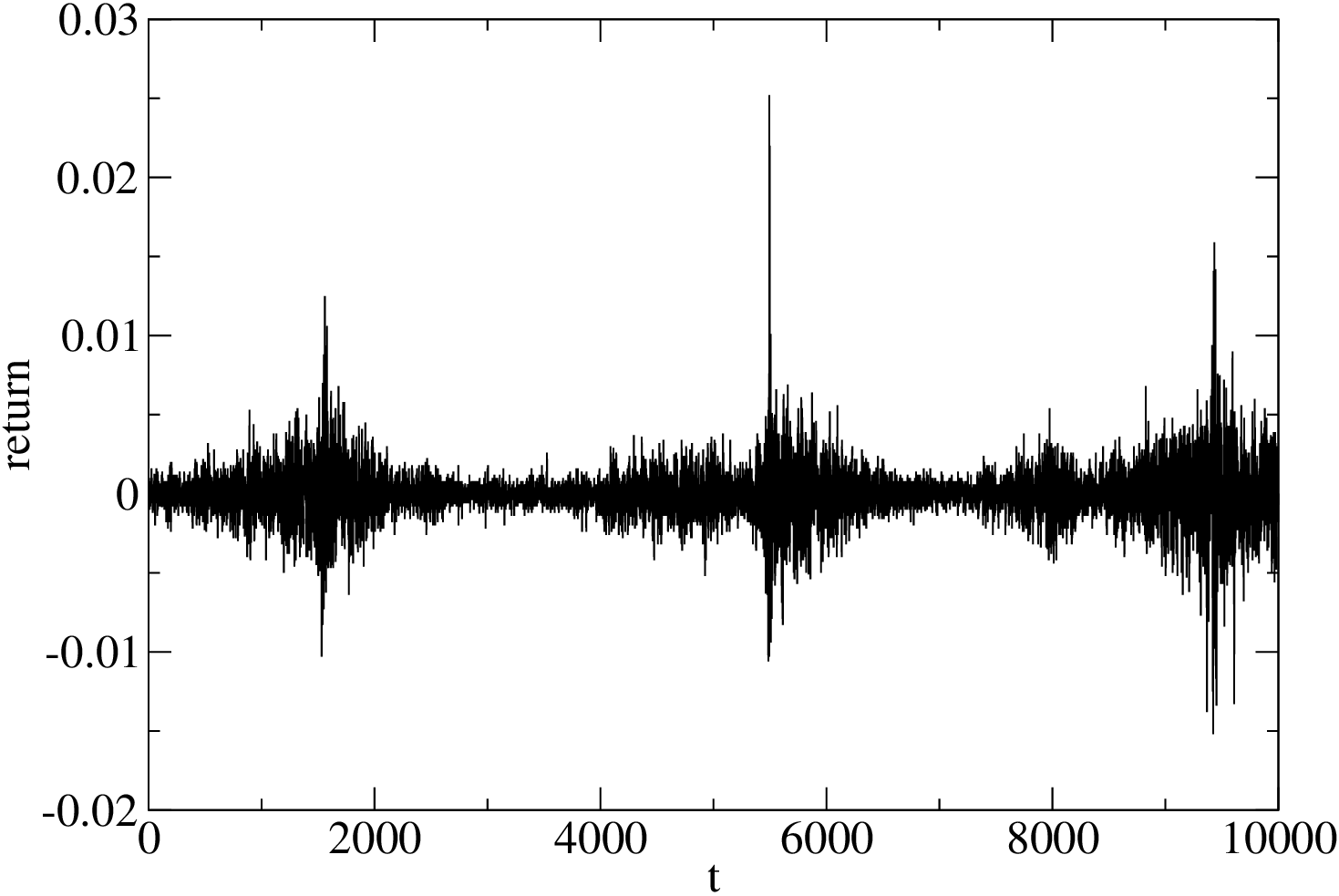}
\vspace{5mm}
\includegraphics[width=17pc]{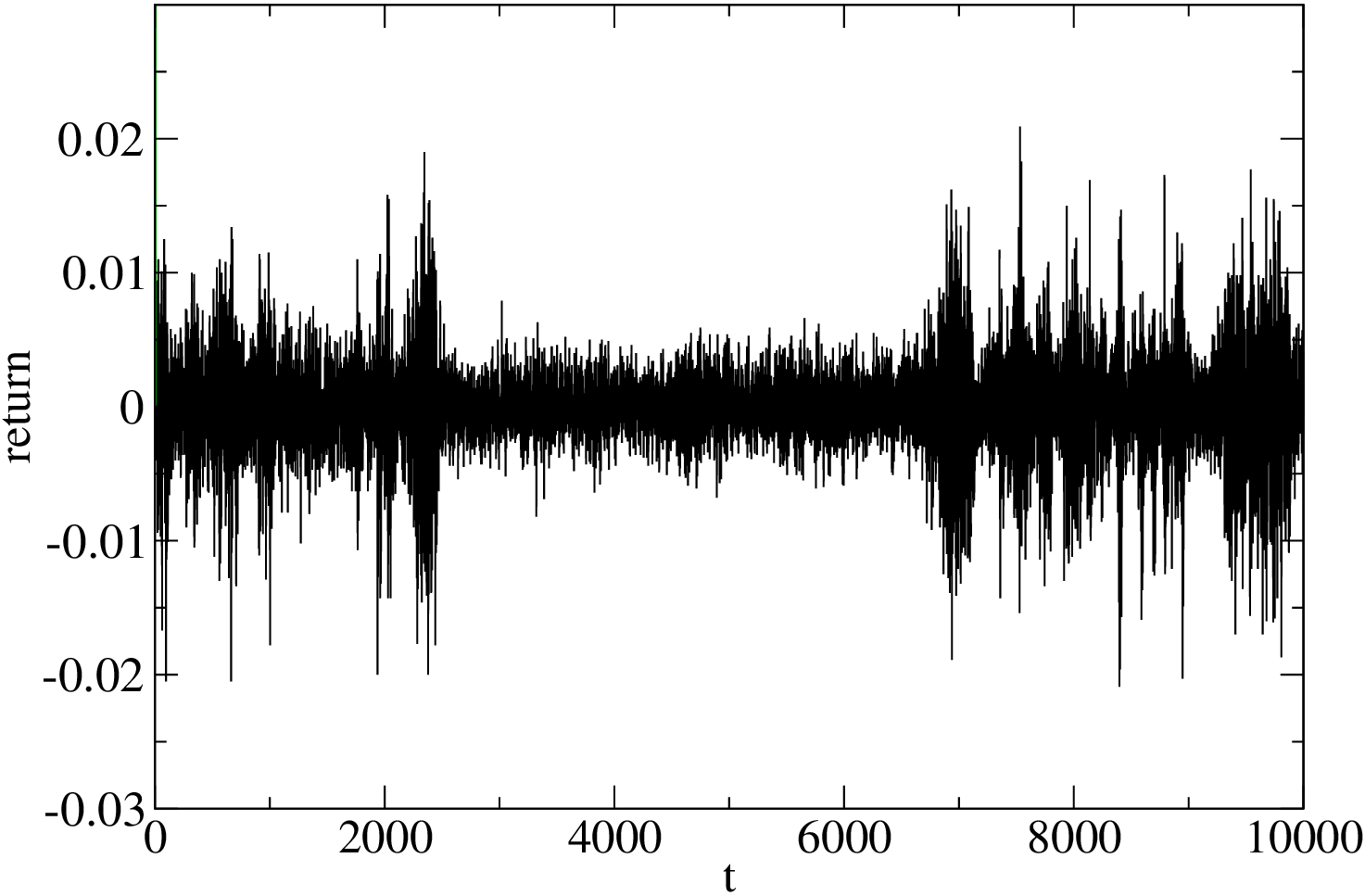}
\end{center}
\caption{\label{label2}Return time series simulated at $\gamma=1.0$ (left) and $\gamma=0.05$ (right).
}
%\end{minipage}
%\hspace{5pc}%
\end{figure}

\section{Simulations of three-state spin model}

We use a $100 \times 100$ square lattice with the periodic boundary condition
and start simulations on a random configuration.
The spins are updated one by one in a random order.
We define "one sweep" as $100^2$ updates and use one sweep as unit time.
We discard the first $3\times 10^4$ sweeps as thermalization and then accumulate $10^4$ sweeps 
for analysis.
We make simulations for two parameter sets (a):$(\lambda ,\mu,\gamma)=(4.0,10.0,1.0)$ and 
(b):$(\lambda ,\mu,\gamma)=(4.0,10.0,0.05)$, i.e. $\lambda$  and $\mu$ are fixed 
and $\gamma$ is varied. 
It is found that the power-law return distribution is obtained for $\gamma=1.0$
and the exponential return distribution for $\gamma=0.05$\cite{TakaishiSP}.

Figure 1 shows the magnetization $M(t)$ as a function of time ( sweep ).
Following \cite{SPIN5} we define the return $r(t)$ through the magnetization as
\be
r(t)=M(t+1)-M(t),
\ee
where $M(t)$ is the magnetization given by (\ref{eq:M}).
Figure 2 shows time series of return $r(t)$ where 
we observe the intermittency of returns.  
As mentioned in section 2
such intermittent behavior is expected to be caused by the phase change.
Namely the ordered phase where $|M|$ and its fluctuation are small 
corresponds to low-volatility phase. Conversely the disordered phase corresponds to high-volatility phase.
These properties can also be confirmed by comparing figures 1 and 2.
Since values of volatility are not obtained directly from returns  
we estimate them by the GARCH model as used in the volatility estimation in empirical finance.

\begin{figure}[ht]
\vspace{1cm}
%\begin{miniprre}{14pc}
\begin{center}
\includegraphics[width=17pc]{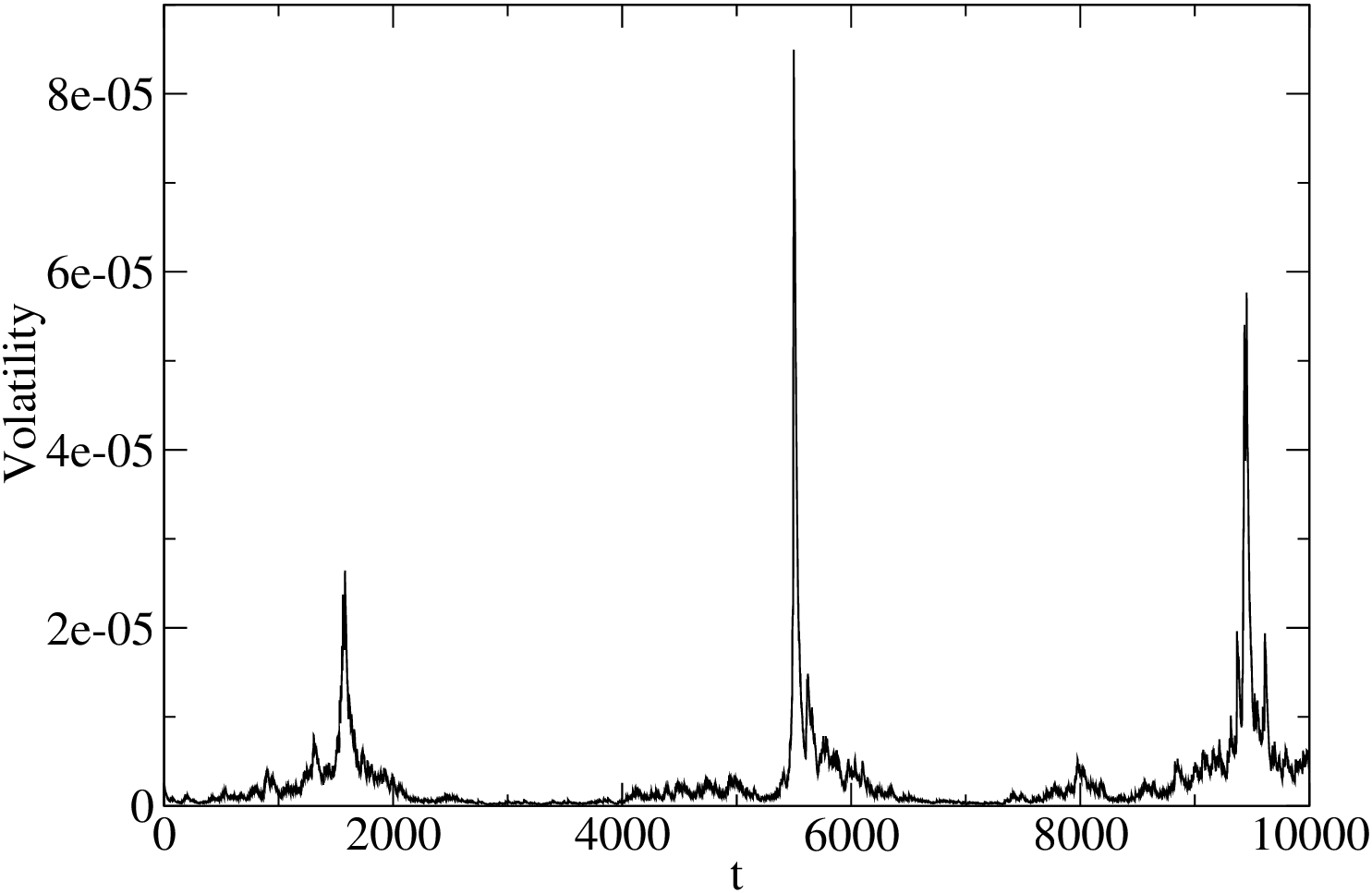}
\hspace{2mm}
\includegraphics[width=17pc]{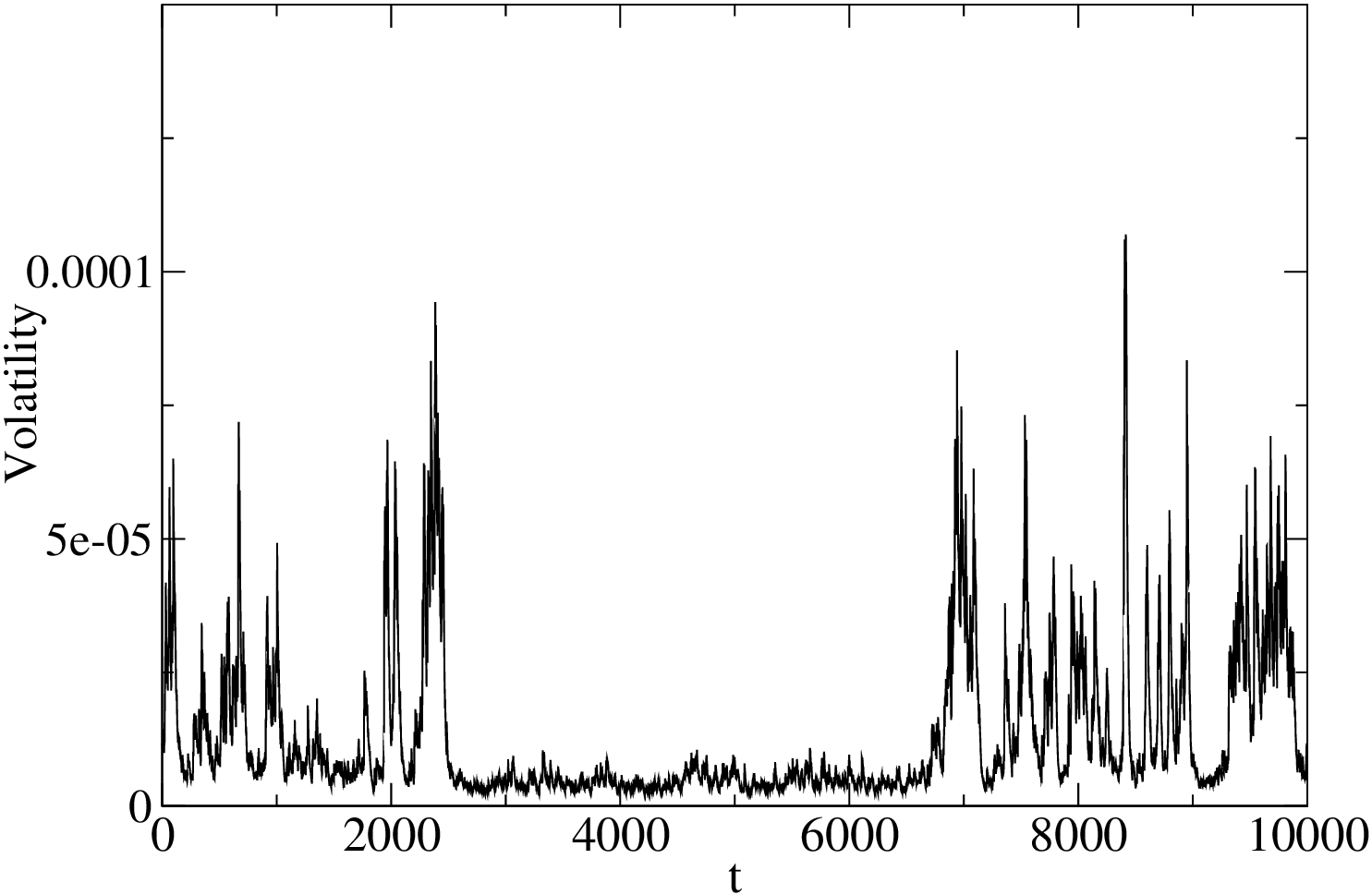}
\end{center}
\caption{\label{label3} Volatility time series obtained by the GARCH model:
$\gamma=1.0$ (left) and $\gamma=0.05$ (right).}
%\end{minipage}
%\hspace{5pc}%
\end{figure}

%\begin{figure}[h]
%\vspace{1cm}
%%\begin{miniprre}{14pc}
%\begin{center}
%\includegraphics[width=16.9pc]{b4a10c1-st-return.eps}
%\hspace{2mm}
%\includegraphics[width=17pc]{b4a10c0.05-st-return.eps}
%\end{center}
%\caption{\label{label4}Standardized return time series simulated
%$\gamma=1.0$ (left) and $\gamma=0.05$ (right).}
%%\end{minipage}
%%\hspace{5pc}%
%\end{figure}

\begin{figure}[ht]
\vspace{1cm}
%\begin{miniprre}{14pc}
\begin{center}
\includegraphics[width=17pc]{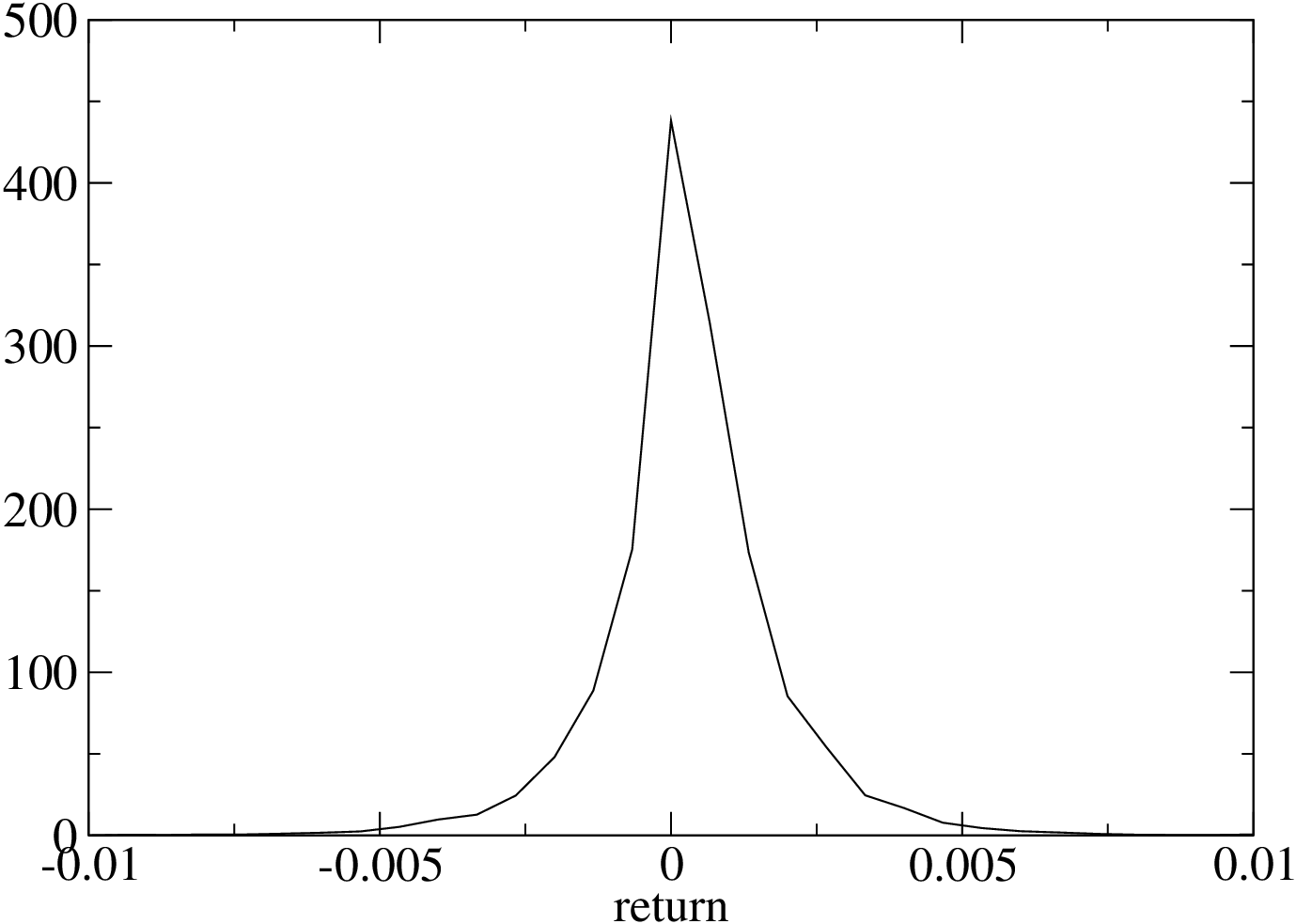}
\hspace{5mm}
\includegraphics[width=16.2pc]{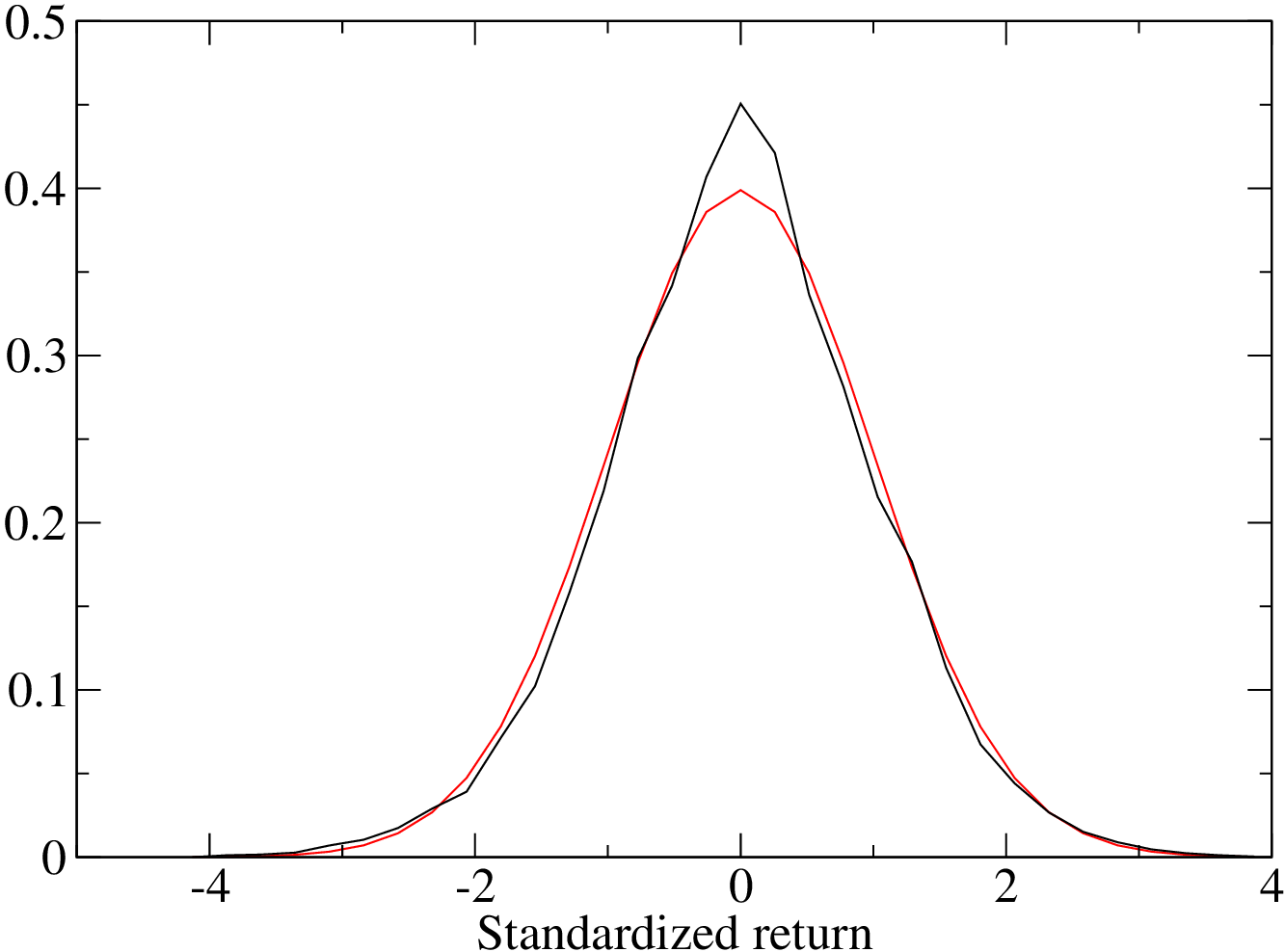}
\end{center}
\caption{\label{label5}Return distribution (left) and standardized return distribution (right)
simulated at $\gamma=1.0$. The red line shows the standard normal distribution.}
%\end{minipage}
%\hspace{5pc}%
\end{figure}

\begin{figure}[ht]
\vspace{1cm}
%\begin{miniprre}{14pc}
\begin{center}
\includegraphics[width=16.8pc]{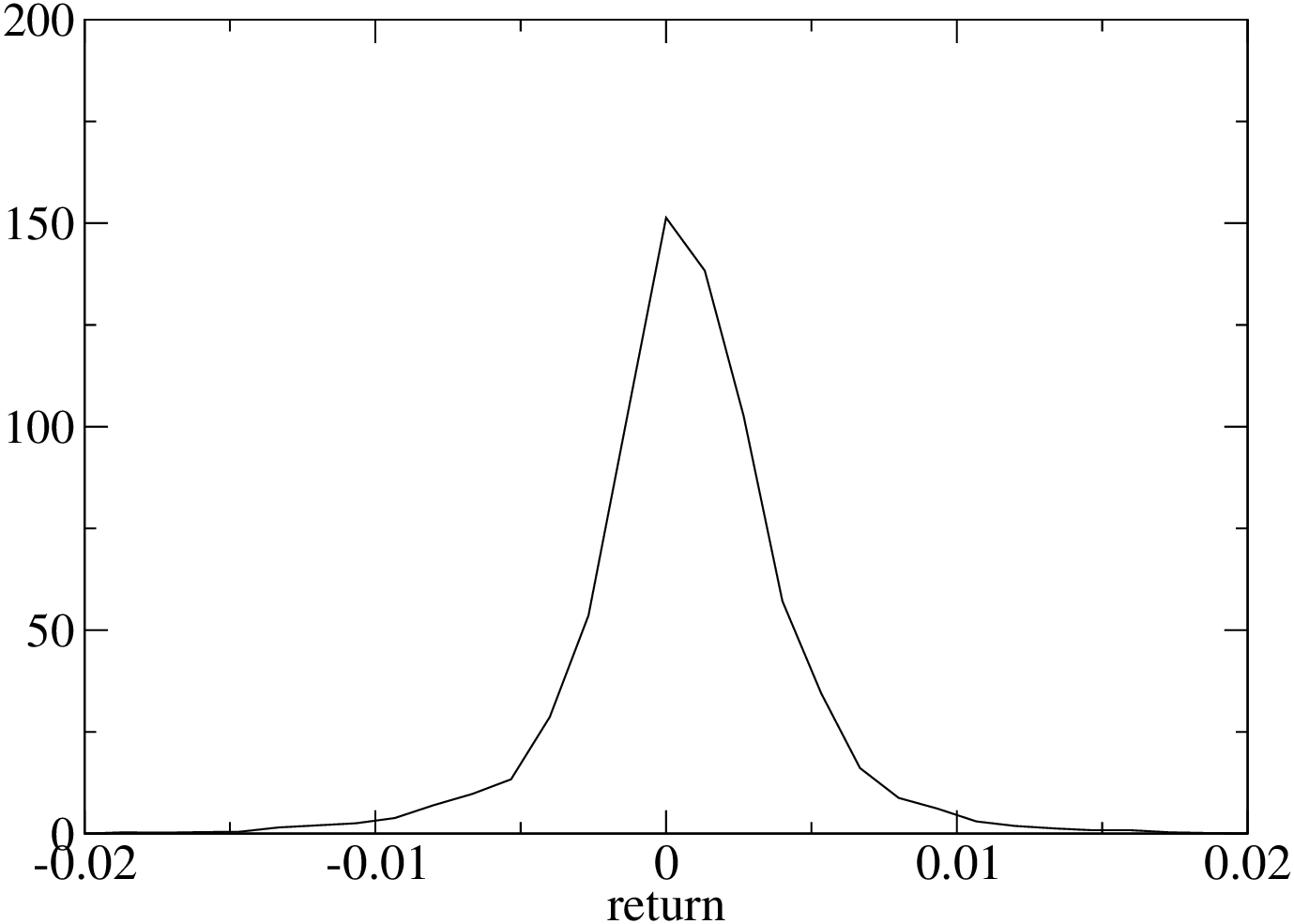}
\hspace{7mm}
\includegraphics[width=16.2pc]{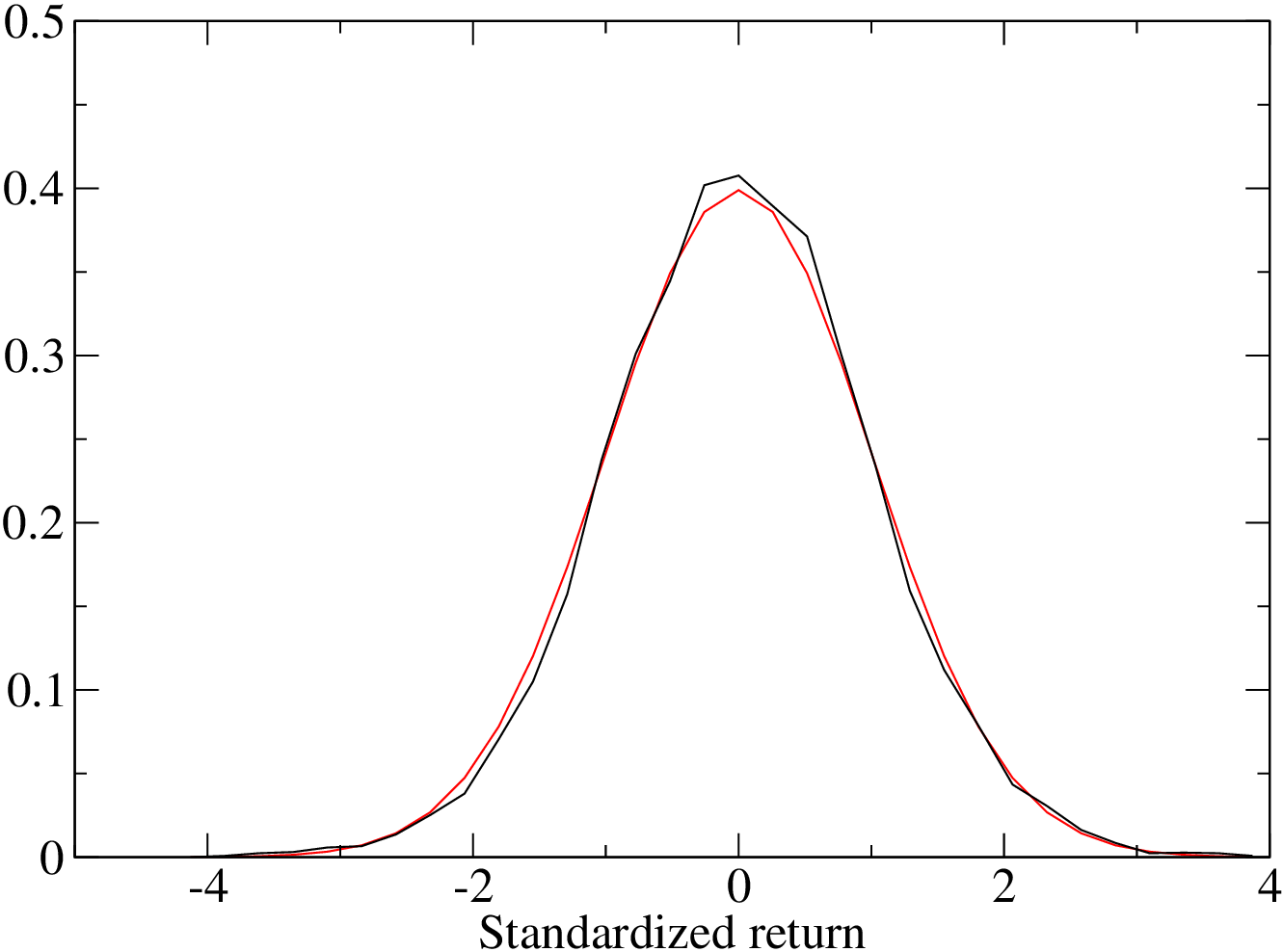}
\end{center}
\caption{\label{label6}
Return distribution (left) and standardized return distribution (right)
simulated at $\gamma=0.05$. The red line shows the standard normal distribution.}
%\end{minipage}
%\hspace{5pc}%
\end{figure}

\section{Volatility from the GARCH model}

We apply the GARCH model for return data obtained from the financial spin model
and estimate volatility corresponding to each return.
The Bayesian inference of the GARCH model is performed by
the Metropolis-Hastings algorithm with 
an adaptive multi-dimensional student's t-distributions\cite{Takaishi1,Takaishi2,Takaishi3,Takaishi4}.
In total we make $4\times 10^4$  Monte Carlo updates of GARCH parameters.
The first $1.1 \times 10^4$ updates are discarded as thermalization and 
the $2.9\times 10^4$ updates are used for analysis.
At each update we also store the values of volatility and 
at the end we average the values of volatility for final output. 
The GARCH parameters determined by the Bayesian inference
are listed in table 1.
For both simulation parameters, values of $\alpha+\beta$ are found to be very close to 1, 
which means that the return time series obtained from those simulations 
have the strong persistency in volatility.

\begin{table}[ht]
  \centering
  \caption{Results of GARCH parameters.
   SD and SE stand for standard deviation and statistical error respectively.
Statistical errors are estimated by the jackknife method. $\tau_{int}$ is the autocorrelation time
defined by $\tau_{int}=1 +2\sum_{i=1}^{\infty}ACF(t)$ where $ACF(t)$ stands for the autocorrelation function.}
  \label{tab:1}
    \begin{tabular}{clll}
      \hline
      & \multicolumn{1}{c}{$\alpha$} &
      \multicolumn{1}{c}{$\beta$} &
      \multicolumn{1}{c}{$\omega$} \\
\hline
   $\gamma=1.0$ & 0.0495  & 0.9508  &  $2.1 \times 10^{-9}$ \\ 
   SD           & 0.0037  & 0.0013  &  $6 \times 10^{-10}$   \\  
   SE           & 0.00003 & 0.00003 &  $4 \times 10^{-12}$ \\
   $\tau_{int}$    & $1.8 \pm 0.4$    & $1.8 \pm 0.3$ & $1.8\pm 0.4$   \\
\hline
   $\gamma=0.05$  & 0.0836  & 0.9100  & $8.6 \times 10^{-8}$   \\
   SD             & 0.0055  & 0.0057  & $1.4  \times 10^{-8}$  \\
   SE             & 0.00003 & 0.00003 & $8 \times 10^{-11}$  \\
   $\tau_{int}$    & $1.41 \pm 0.05$  & $1.40\pm 0.04$  & $1.42\pm 0.05$  \\
\hline
    \end{tabular}
\end{table}

Figure 3 shows the time series of volatility obtained by the GARCH model.
It is seen that there exist high-volatility and low-volatility periods.
Such behavior of the volatility corresponds to "volatility clustering" which 
often observed in the real financial markets.

Next we examine the view of the MDH for the return data from the financial spin model.
In the MDH the return $r(t)$ is assumed to be given by 
$r(t)=\sigma_t \epsilon_t$,
where $\sigma_t^2$ is volatility and 
$\epsilon_t$ is an independent Gaussian random variable with mean 0 and variance 1.
Under this assumption 
the returns standardized by $\sigma_t$, i.e. $r(t)/\sigma_t$
should be standard normal random variables.
Using volatility determined by the GARCH model as a proxy of the true volatility
we standardize returns $r(t)$. 

Figures 4 and 5 compare the distributions of the unstandardized and standardized returns.
The unstandardized return distributions are very different from 
the standard normal distributions ( variance 1 and mean 0 ).
On the other hand the standardized return distributions come close to
the standard normal distributions that are  depicted in red.
The  normality of the  standardized returns can be further confirmed by 
examining variance and kurtosis.

Table 2 lists variance and kurtosis of the unstandardized and standardized returns.
The variance and kurtosis of the unstandardized returns differ from 
the values expected for standard normal random variables, i.e. variance = 1 and kurtosis = 3.
On the other hand it is remarkable that the variances of standardized returns from both simulations 
at $\gamma = 1.0$ and 0.05 are consistent with 1.
The values of kurtosis for standardized returns also come
close to 3 although their values are slightly higher than 3.
This small disagreement might be understood by that
the GARCH model we used here, i.e. the GARCH(1,1) model with normal errors,  
still does not capture completely the properties of the return data from
the spin financial model.
The similar disagreement has been observed in 
the real financial markets\cite{Andersen4}.

We further examine the MDH by calculating the autocorrelation function.
As a stylized fact of asset returns
it is known that  the absolute return time series exhibits a very long correlation.
However under the MDH the standardized returns are expected to be 
Gaussian random variables. 
Therefore if the MDH is hold we expect that
the autocorrelation between the absolute standardized returns disappears. 
Figure 6 shows the autocorrelation functions of the absolute returns and absolute standardized returns.
It is clear that  the absolute returns show a very long correlation but 
the absolute standardized returns have no significant correlation.
Thus this examination of the autocorrelation function also confirms 
the MDH for the return time series of the spin financial market.
The similar behavior of the autocorrelation function on the absolute standardized returns 
is also observed in the real financial market\cite{TakaishiRV}. 

\begin{table}
\caption{Variance and kurtosis of the unstandardized and standardized returns.}
\begin{tabular}{c|cc|cc} \hline
               & \multicolumn{2}{c}{$\gamma=1.0$} &  \multicolumn{2}{c}{$\gamma=0.05$} \\ \hline
               & variance  & kurtosis &variance & kurtosis \\
unstandardized & $2.7 \times 10^{-6} \pm 9 \times 10^{-7}$ & $21 \pm 9$ & $1.3 \times 10^{-5} \pm 2 \times 10^{-6}$ & $6.6 \pm 0.8$ \\
standardized   & $ 0.997  \pm 0.022 $                         & $3.54 \pm 0.09$ & $0.999 \pm 0.025$ & $3.49 \pm 0.10 $ \\
\hline
\end{tabular}

\end{table}

\begin{figure}[ht]
\vspace{1cm}
%\begin{miniprre}{14pc}
\begin{center}
\includegraphics[width=17pc]{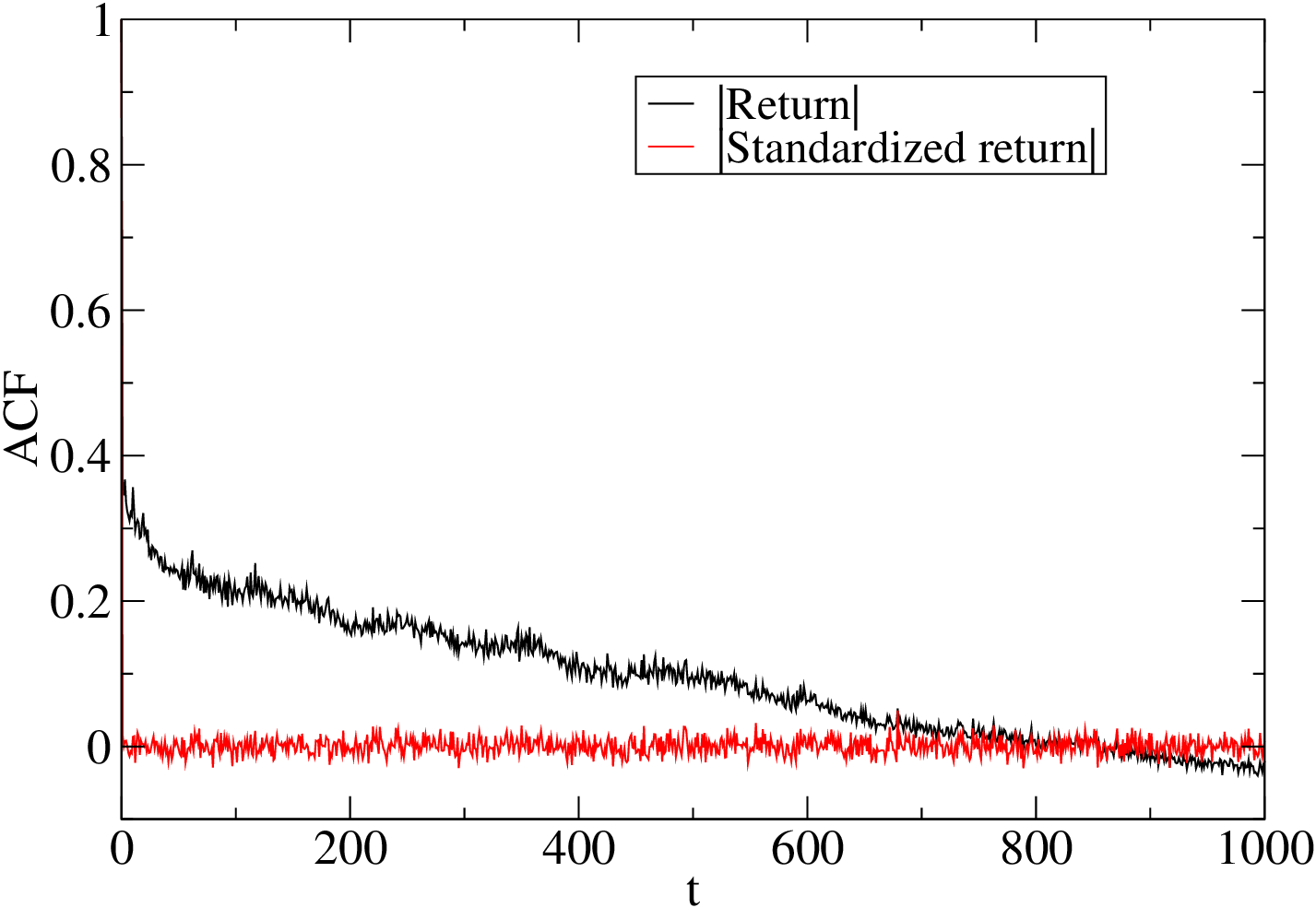}
\hspace{2mm}
\includegraphics[width=17pc]{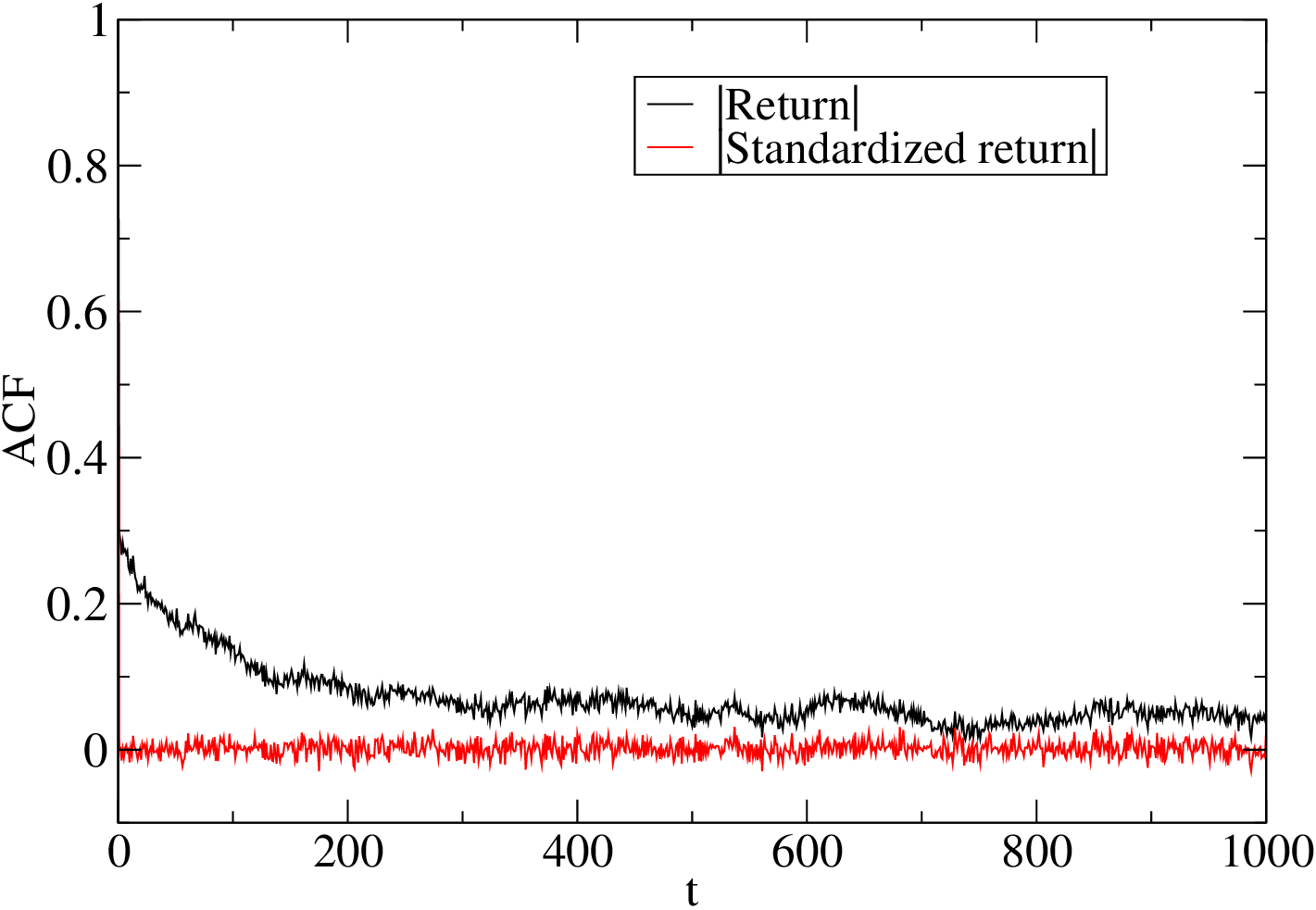}
\end{center}
\caption{Autocorrelation functions of absolute returns and absolute standardized returns
at $\gamma=1.0$ (left) and $\gamma=0.05$ (right).}
%\end{minipage}
%\hspace{5pc}%
\end{figure}

\section{Conclusions}
The volatility of a financial spin model with three-states was investigated by 
the GARCH model.
To determine volatility by the GARCH model 
the Bayesian inference by the MCMC method was applied.
We find that volatility of the spin financial market
exhibits "volatility clustering" which is often observed 
in the real financial markets.
In order to examine the MDH for the spin financial market we analyzed returns 
standardized by the volatility obtained from the GARCH model.
Under the MDH the normality is expected for the standardized returns.
We find that variance and kurtosis of the standardized returns
are very similar to those of standard normal random variables
except for a slightly higher kurtosis. 
The autocorrelations of the absolute standardized returns were
also investigated and no significant autocorrelation between  absolute standardized returns 
was found, which can also be expected from the MDH.
From these findings it is concluded that 
the spin financial market we simulated here is  consistent with
the view of the MDH for the return dynamics.

\section*{Acknowledgemet}
Numerical calculations in this work were carried out at the
Yukawa Institute Computer Facility
and the facilities of the Institute of Statistical Mathematics.
This work was supported by Grant-in-Aid for Scientific Research (C) (No.22500267).


\section*{References}
\begin{thebibliography}{9}
%\bibitem{iopartnum} IOP Publishing is to grateful Mark A Caprio, Center for Theoretical Physics, Yale University, for permission to include the {\tt iopart-num} \BibTeX package (version 2.0, December 21, 2006) with  this documentation. Updates and new releases of {\tt iopart-num} can be found on \verb"www.ctan.org" (CTAN). 
\bibitem{CONT}
%R.~Cont, Empirical Properties of Asset Returns: Stylized Facts and Statistical Issues,
%{\it Quantitative Finance} {\bf 1} (2001) 223--236.
Cont R 2001
Empirical Properties of Asset Returns: Stylized Facts and Statistical Issues
{\it Quantitative Finance} {\bf 1}  223--236

\bibitem{Clark}
Clark  P K 1973 
A subordinated stochastic process model with finite variance for speculative prices  
{\it Econometrica} {\bf 41}  135-155


%\bibitem[(Andersen {\it et al.}, 2000)]{Andersen0}
\bibitem{Andersen1}
%T.G.~Andersen, T.Bollerslev, F.X.Diebold and P.Labys (2000)
Andersen T G,  Bollerslev T,  Diebold F X and Labys P 2000
Exchange Rate Returns Standardized by Realized Volatility are (Nealy) Gaussian
\textit{Multinational Finance Journal} {\bf 4}  159--179


\bibitem{Andersen2}
%T.G.~Andersen, T.Bollerslev, F.X.Diebold and H.Ebens,
%\textquotedblleft The distribution of realized stock return volatility,"
%\textit{Journal of Financial Economics 61} (2001), 43--76.
Andersen T G,  Bollerslev T,  Diebold F X and Ebens H 2001
The distribution of realized stock return volatility
\textit{Journal of Financial Economics} {\bf 61}  43--76

\bibitem{Andersen3}
%Andersen, T. G.,   Bollerslev, T., & Dobrev, D. (2007). 
Andersen T G,  Bollerslev T and Dobrev, D 2007
No-arbitrage semi-martingale restrictions for
continuous-time volatility models subject to leverage effects, jumps and i.i.d. noise: Theory and testable distributional implications
{\it Journal of Econometrics}  {\bf 138}  125-180

\bibitem{Andersen4}
Andersen T G,   Bollerslev T, Frederiksen P and Nielsen M {\O} 2010
Continuous－time models, realized volatilities, and testable distributional implications for daily stock returns 
{\it Journal of Applied Econometrics} {\bf 25}  233-261

\bibitem{Fleming}
Fleming J and Paye B S 2011 
High-frequency returns, jumps and the mixture of normals hypothesis
{\it Journal of Econometrics} {\bf 160} 119-128

\bibitem{TakaishiRV}
Takaishi T, Chen TT and Zheng Z 2012
Analysis of Realized Volatility in Two Trading Sessions of the Japanese Stock Market
\textit{Prog. Theor. Phys. Supplement} {\bf 194}  43-54

\bibitem{TakaishiRV2}
Takaishi T 2012
Finite-Sample Effects on the Standardized Returns of the Tokyo Stock Exchange 
{\it Procedia - Social and Behavioral Sciences} {\bf 65} 968--973

\bibitem{ARCH}
Engle  R  F  1982
Autoregressive Conditional Heteroskedasticity with Estimates of the Variance of the United Kingdom inflation 
{\it Econometrica} {\bf 50}  987--1007

\bibitem{GARCH}
Bollerslev  T 1986  
Generalized Autoregressive Conditional Heteroskedasticity 
{\it Journal of Econometrics} {\bf 31}  307--327

\bibitem{EGARCH}
Nelson D B 1991 Conditional Heteroskedasticity in Asset Returns: A New Approach
{\it Econometrica} {\bf 59}  347-370
%Econometrica 59, 347--370 (1991)

\bibitem{QGARCH1}
%R.F.~Engle, V.~Ng,
Engle  R F and  Ng  V 1993
Measuring and testing the impact of news on volatility
{\it Journal of Finance} {\bf 48}  1749-1778
%Journal of Finance 48, 1749--1778 (1993)

\bibitem{QGARCH2}
%E.~Sentana,
Sentana  E 1995
Quadratic ARCH models
{\it Review of Economic Studies} {\bf 62}  639-661
%Review of Economic Studies 62, 639--661 (1995)

\bibitem{GJR}
%L.R.~Glosten, R.~Jaganathan, D.E.~Runkle,
Glosten L R, Jaganathan  R and  Runkle  D E 1993
On the Relation Between the Expected Value and the Volatility of the Nominal Excess on Stocks
{\it Journal of Finance} {\bf 48}  1779-1801
%Journal of Finance 48, 1779--1801 (1993)

\bibitem{APARCH}
Ding Z, Granger C W J and Engle R F 1993
A long memory property of stock market returns and a new model
{\it Journal of Empirical Finance} {\bf 1} 83--106

\bibitem{GARCHRE}
Takaishi T and Chen TT 2012
Bayesian Inference of the GARCH model with Rational Errors
{\it International Proceedings of  Economics Development and Research} {\bf 29} 303--307


\bibitem{Bollerslev}
Bollerslev  T, Chou R Y and Kroner K  F 1992
ARCH modeling in finance
{\it Journal of Econometrics} {\bf 52} 5--59

%Agent model
\bibitem{Bouchaud}
Cont R and Bouchaud J P 2000
Herd behaviour and aggregate fluctuations in financial markets
{\it Macroecon. Dynamics} {\bf 4} 170-196

\bibitem{Stauffer}
Stauffer D and Penna T J P 1998
Crossover in the Cont-Bouchaud percolation model for market fluctuations 
{\it Physica A} {\bf 256} 284--290

\bibitem{Sato} 
Sato A H and Takayasu H 1998
Dynamic numerical models of stock market price: from microscopic determinism ot macroscopic randomness
{\it Physica A} {\bf 250} 231--252

\bibitem{Lux}
Lux T and Marchesi M 1999
Scaling and Criticality in a Stochastic Multi-Agent Model of a Financial Market
{\it Nature} {\bf 397}  498--500

\bibitem{Iori}
%G. Iori,
%Avalanche Dynamics and Trading Friction Effects on Stock Market Returns,
%{\it Int. J. Mod. Phys. C} {\bf 10} (1999) 1149--1162.
Iori G 1999
Avalanche Dynamics and Trading Friction Effects on Stock Market Returns
{\it Int. J. Mod. Phys. C} {\bf 10}  1149--1162

\bibitem{Stauffer2}
%L.R. da Silva and D. Stauffer,
%Ising-correlated clusters in the Cont-Bouchaud stock market model,
%{\it Physics A} {\bf 294} (2001) 235--238.
da Silva L R  and Stauffer D 2001
Ising-correlated clusters in the Cont-Bouchaud stock market model
{\it Physics A} {\bf 294} 235--238


%\bibitem{Zhang}
%D. Challet, A. Chessa, M. Marsili and Y-C. Zhang, From Minority Games to real markets,
%{\it Quantitative Finance} {\bf 1} (2001) 168--176.
%(2001)

\bibitem{AGENT}
Challet D, Marsili M and Zecchina R 2000
Statistical mechanics of system with heterogeneous agents: Minority games
{\it Phys. Rev. Lett. } {\bf 84(8)} 1824--1827


\bibitem{AGENT2}
%M. Raberto, S. Cincotti, S.M. Focardi and M. Marchesi,
%Agent-based Simulation of a Financial Market,
%{\it Physics A} {\bf 299} (2001) 319--327.
Raberto M, Cincotti S, Focardi S M and  Marchesi M 2001
Agent-based Simulation of a Financial Market
{\it Physics A} {\bf 299}  319--327

%SPIN MODEL
\bibitem{Bornholdt}
%S.~Bornholdt,
%Expectation Bubbles in a Spin Model of Markets: Intermittency from Frustration across Scales.
%%EXPECTATION BUBBLES IN A SPIN MODEL OF MARKETS: INTERMITTENCY FROM FRUSTRATION ACROSS SCALES
%{\it Int. J. Mod. Phys. C} {\bf 12} (2001) 667--674.
Bornholdt S 2001
Expectation Bubbles in a Spin Model of Markets: Intermittency from Frustration across Scales
%EXPECTATION BUBBLES IN A SPIN MODEL OF MARKETS: INTERMITTENCY FROM FRUSTRATION ACROSS SCALES
{\it Int. J. Mod. Phys. C} {\bf 12}  667--674

\bibitem{SPIN1}
%K.~Sznajd-Weron and R.~Weron,
%A Simple Model of Price Formation.
%%A SIMPLE MODEL OF PRICE FORMATION
%{\it Int. J. Mod. Phys. C} {\bf 13} (2002) 115--123.
Sznajd-Weron K and Weron R 2002 
A Simple Model of Price Formation
%A SIMPLE MODEL OF PRICE FORMATION
{\it Int. J. Mod. Phys. C} {\bf 13}  115--123

\bibitem{Sanchez}
Sanchez J R 2002
A Simple Model for  Stocks Markets
{\it Int. J. Mod. Phys. C} {\bf 13}  639--644

\bibitem{Yamano}
%T. Yamano,
%Bornholdt's Spin Model of a Market Dynamics in High Dimensions,
%{\it Int. J. Mod. Phys. C} {\bf 13} (2002) 89--96.
Yamano Y 2002
Bornholdt's Spin Model of a Market Dynamics in High Dimensions
{\it Int. J. Mod. Phys. C} {\bf 13} 89--96

\bibitem{SPIN5}
%Kaizoji, T. {\it et al.}
%T.~Kaizoji, S.~Bornholdt and Y.~Fujiwara,
Kaizoji T, Bornholdt S and Fujiwara Y 2002
Dynamics of Price and Trading Volume in a Spin Model of Stock Markets with Heterogeneous Agents
{\it Physica A} {\bf 316}  441--452

\bibitem{TakaishiSP}
Takaishi T 2005
Simulations of Financial Markets in a Potts-like Model 
{\it Int. J. Mod. Phys. C} {\bf 16}  1311--1317


\bibitem{Bornholdt2}
Krause S M and Bornholdt S 2011
Spin models as microfoundation of macroscopic financial market models
{\it arXiv:1103.5345}

\bibitem{MFA}
Stauffer D 2012
A Biased Review of Sociophysics
{\it J. Stat. Phys.} {\bf 151} 9--20

\bibitem{Indian}
Matia K, Pal M, Salunkay H and Stanley H E 2004
Scale-dependent price fluctuations for the Indian stock market
{\it Europhys. Lett.} {\bf 66} 909--914


\bibitem{AIC}
Akaike H 1973 
Information theory and an extension of the maximum likelihood principle
{\it Proceedings of the 2nd International Symposium on Information Theory}  Petrov B N and Caski F (eds.), Akadimiai Kiado, Budapest: 267-281 


\bibitem{Takaishi1}
%T.~Takaishi,
Takaishi  T 2009
An Adaptive Markov Chain Monte Carlo Method for GARCH Model
{\it Lecture Notes of the Institute for Computer Sciences, Social Informatics and Telecommunications Engineering.
Complex Sciences} Vol. {\bf 5} 1424--1434
%doi:10.1007/978-3-642-02469-6\_22
%{\it Preprint} arXiv:0901.0992v1

\bibitem{Takaishi2}
%T.~Takaishi,
Takaishi T 2009
Bayesian Estimation of GARCH Model with an Adaptive Proposal Density
{\it New Advances in Intelligent Decision Technologies,
Studies in Computational Intelligence} Vol. {\bf 199} 635--643
%Studies in Computational Intelligence} Volume {\bf 199} 635--643 (2009) \\
%doi:10.1007/978-3-642-00909-9\_61

\bibitem{Takaishi3}
%T.~Takaishi,
Takaishi  T 2009
Bayesian Inference on QGARCH Model Using the Adaptive Construction Scheme
{\it Proceedings of 8th IEEE/ACIS International Conference on Computer and Information Science}
525--529
doi:10.1109/ICIS.2009.173

\bibitem{Takaishi4}
Takaishi T 2010
Bayesian inference with an adaptive proposal density for GARCH models
{\it J. Phys.: Conf. Ser.} {\bf 221} 012011


\end{thebibliography}
\end{document}